\documentclass[12pt]{article}
\usepackage{latexsym}
\usepackage{graphicx}
\usepackage{multirow}
\usepackage{amsmath}
\usepackage{ulem}
\usepackage{setspace}
\usepackage{subfigure}
\usepackage{color}
\usepackage{soul}
\begin{document}
\begin{center}\Large{Relations between the baryon quantum numbers of the Standard Model and of the rotating neutrino model}
\end{center}
\begin{center}
\renewcommand\thefootnote{\fnsymbol{footnote}}
C.G. Vayenas$^{1,2,*}$, A.S. Fokas$^{3,4,}$\footnote{E-mail: cgvayenas@upatras.gr; T.Fokas@damtp.cam.ac.uk} \& D.P. Grigoriou$^{1}$
\end{center}
\begin{center}
$^1${School of Engineering, University of Patras, GR 26504 Patras, Greece}\\
$^2${Division of Natural Sciences, Academy of Athens, 28 Panepistimiou Ave., \\ GR-10679 Athens, Greece}\\ $^3${Department of Applied Mathematics and Theoretical Physics, University of Cambridge, Cambridge, CB3 0WA, UK}\\$^{4}${Viterbi School of Engineering, University of Southern California, Los Angeles, California, 90089-2560, USA}
\end{center}

\begin{abstract}
{We discuss the common features between the Standard Model taxonomy of particles, based on electric charge, strangeness and isospin, and the taxonomy emerging from the key structural elements of the rotating neutrino model, which describes baryons as bound states formed by three highly relativistic electrically polarized neutrinos forming a symmetric ring rotating around a central electrically charged or polarized lepton. It is shown that the two taxonomies are fully compatible with each other.} 
\end{abstract}
\vspace{0.5cm}
\vspace{0.5cm}
\section {Introduction}
After it was established experimentally that baryons contain three quarks there were intense efforts to identify a satisfactory taxonomy for a variety of baryons. This goal was satisfactorily achieved in the seventies \cite{Griffiths08,Tully11}, by employing electric charge as well as the concepts of isospin and strangeness, see Figures 1 and 2 \cite{Griffiths08,Tully11}. 
\begin{figure}[t]
\begin{center}
\includegraphics[width=0.45\textwidth]{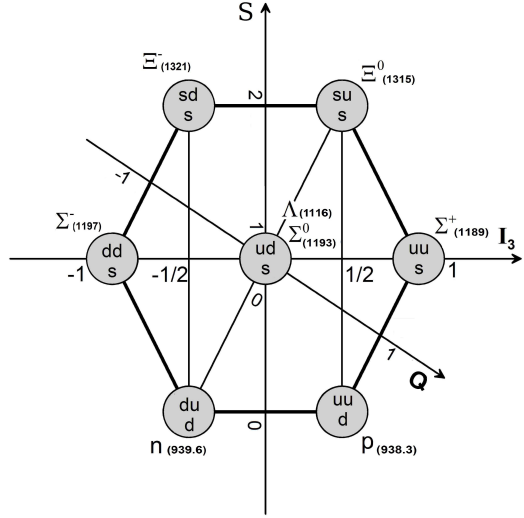}
\caption{Combinations of $u$, $d$ or $s$ quarks form baryons with a spin-$1/2$ creating the $uds$ $baryon$ $octet$. Q is the charge in units of $e$, $S$ is the strangeness and I$_3$ is the isospin. For example, the proton has charge 1, isospin 1/2 and strangeness zero \cite{Griffiths08,Tully11}.}
\label{fig:1}
\end{center}
\end{figure}

In several recent works \cite{Vayenas12,Vayenas1106} we have introduced a relativistic neutrino model which treats quarks as electrically polarized neutrinos, see Figure 3. This model does not contain adjustable parameters and allows for the computation of baryon masses within $1\%$ accuracy. It is a Bohr-type model and is based on the use of the de Broglie wavelength equation, together with the Newtonian gravitational law utilizing gravitational rather than rest masses \cite{Vayenas12,Vayenas1106}, namely
\begin{figure}[t]
\begin{center}
\includegraphics[width=0.45\textwidth]{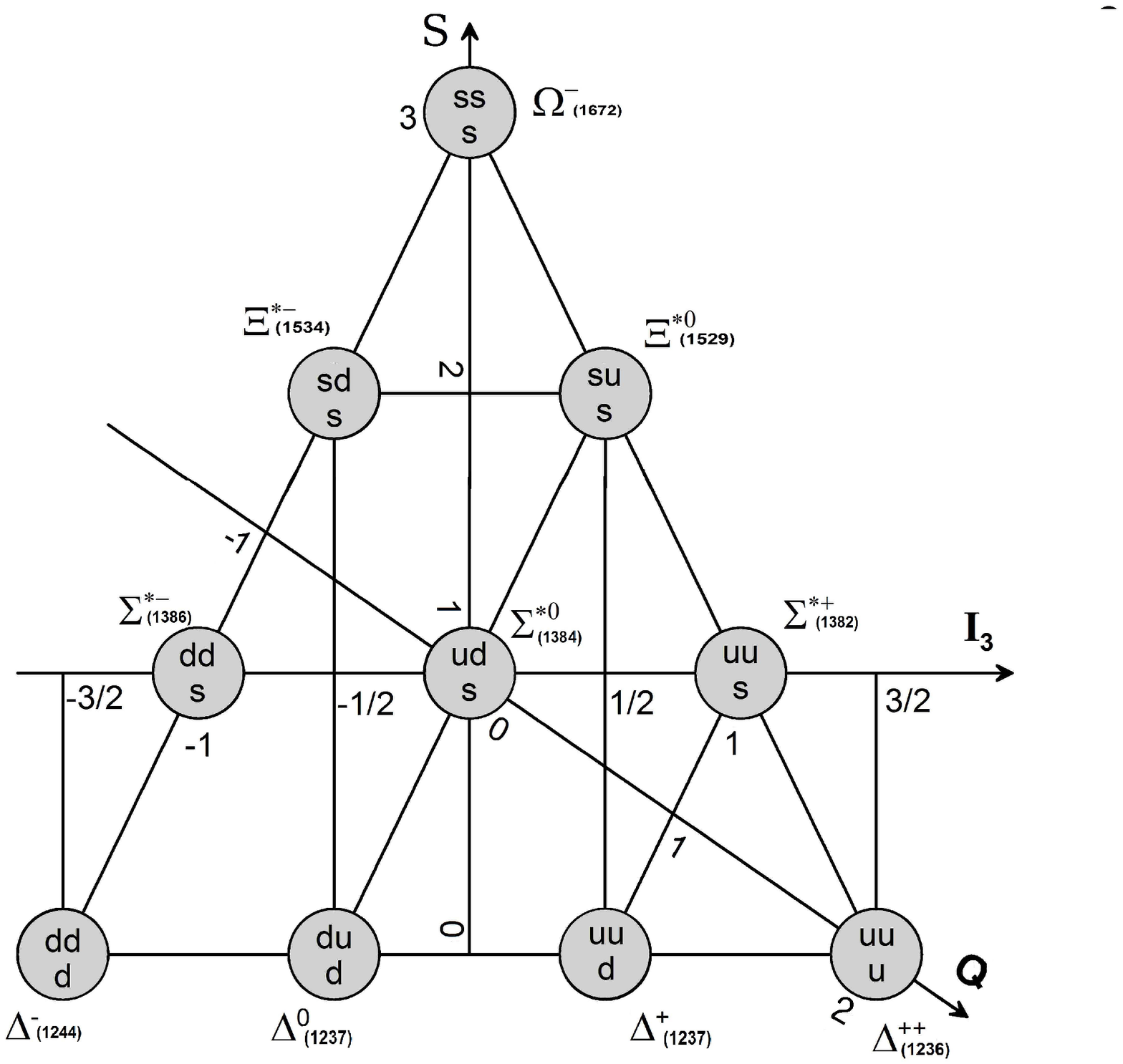}
	\caption{Combinations of $u$, $d$ or $s$ quarks form baryons with a spin-$3/2$ creating the $uds$ $baryon$ $decuplet$ \cite{Griffiths08,Tully11}.}
	\label{fig:2}
	\end{center}
\end{figure}

\begin{equation}
\label{eq1}
F=\frac{Gm_{g,1}m_{g,2}}{\sqrt{3}r^2}=\frac{Gm^2_o\gamma^6}{\sqrt{3}r^2},
\end{equation}
where $m_{g,1}$ and $m_{g,2}$ are the gravitational masses of the two attracting particles which, according to the equivalence principle \cite{Roll1964} are equal to the inertial masses, $m_{i,1}$ and $m_{i,2}$, and which according to Einstein's pioneering special relativity paper \cite{Einstein1905} equal $\gamma^3 m_o$, where $\gamma(=(1-\texttt{v}^2/c^2)^{-1/2})$ is the Lorentz factor and $m_o$ is the rest mass of each particle. This result has been proven by Einstein for linear motion \cite{Einstein1905,French68,Freund08} and was shown recently \cite{Vayenas12,Vayenas1106} that it also holds for arbitrary motion. The factor $\sqrt{3}$ in equation (\ref{eq1}) is due to the triangular geometry. 

Equation (\ref{eq1}) together with the relativistic equation of motion for circular motion 
\begin{equation}
\label{eq2}
F=\gamma m_o\frac{\texttt{v}^2}{r},
\end{equation}
and together with the de Broglie wavelength equation
\begin{equation}
\label{eq3}
\mathchar'26\mkern-10mu\lambda=\frac{\hbar}{\gamma m_o\texttt{v}},
\end{equation}
and the assumption $\mathchar'26\mkern-10mu\lambda=r$, leads to the expression 
\begin{equation}
\label{eq4}
F=\frac{\gamma m_o\texttt{v}^2}{r}=\frac{\hbar \texttt{v}}{r^2}.
\end{equation}
Equations (\ref{eq1}) - (\ref{eq3}) imply \cite{Vayenas12,Vayenas1106} that $\texttt{v}\approx c$ and $\gamma=3^{1/12}(m_{Pl}/m_o)^{1/3}$, where $m_{Pl}(=(\hbar c/G)^{1/2})$ is the Planck mass. Thus the mass, $3\gamma m_o$, of the proton and of the neutron can be approximated by
\begin{equation}
\label{eq5}
m_B=3^{13/12}(m_{Pl}m_o^2)^{1/3}=938.32\;MeV/c^2,
\end{equation}
where we have used $m_o=0.04372\;eV/c^2$ \cite{Vayenas12,Vayenas1106} for the highest neutrino mass \cite{Mohapatra07,Harald}. More recently the approach has been extended to describe the mass of W$^\pm$ \cite{Vayenas16}, Z$^o$ \cite{ICMSQUARE} and H$^o$ bosons, modeled as rotational $e^\pm\nu_e$ structures, for which the following expressions have been obtained:
\begin{equation}
\label{eq6}
m_{W^{\pm}}=2^{1/3}(m_{Pl}m_em_o)^{1/3}=81.74\;GeV/c^2,
\end{equation}
\begin{equation}
\label{eq7}
m_{Z}=2^{1/2}(m_{Pl}m_em_o)^{1/3}=91.75\;GeV/c^2,
\end{equation}
\begin{equation}
\label{eq8}
m_{H}=2^{11/12}(m_{Pl}m_em_o)^{1/3}=122.5\;GeV/c^2.
\end{equation}

In the recent works \cite{Das09,Vayenas15} it is argued that hadrons may be viewed as microscopic black holes \cite{Carr11,Carr14}, where gravitational collapse is prevented by the uncertainty principle. Furthermore, in \cite{Corda13,Corda15} Bohr-type models are used in order to establish a link between black holes and quantum gravity.

The relativistic Newtonian gravitational law (eq. (\ref{eq1})) used to derive the above expressions has also been shown recently to yield exactly the same results as general relativity in the well known problem of the perihelion advancement of Mercury \cite{Fokas15}. 

In the present work we compare the predictions of equation (\ref{eq1}) and of the associated rotating-neutrino-lepton model, with experimental results and with the corresponding description of the Standard Model.
\begin{figure}[t]
\includegraphics[width=0.35\textwidth]{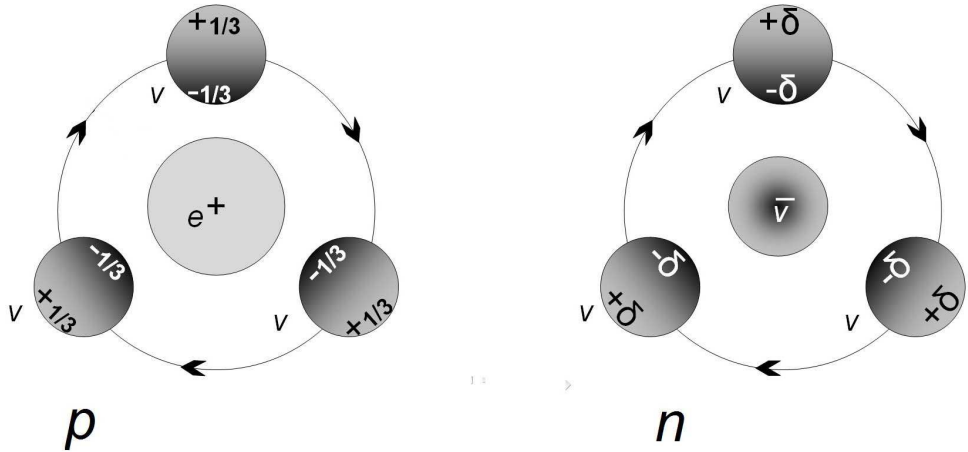}
\caption{Schematics of a proton and a neutron according to the rotating neutrino model, showing the electric dipoles induced on the rotating neutrinos; $\delta$ denotes the induced partial dipole charge \cite{Vayenas16}. Reprinted with permission from Elsevier.}
\label{fig:3}
\end{figure}

\section {Standard Model taxonomy}
The taxonomy of hadrons according to the Standard Model is based on charge, $Q$, isospin, $I$, and strangeness, $S$. This is manifested in the octet of baryons shown in Figure 1, named by Gell-mann the ``eightfold way'' \cite{Gell-Mann53,Nakano53} and the decuplet of baryons \cite{Gell-Mann53,Greenberg82} shown in Figure 2.

Multiplets of the octet and of the decuplet can be obtained by employing higher representations of the symmetry group SU(3), a higher symmetry than the SU(2) of isospin theory. 

One observes in these figures that $I_3$ increases with increasing electrical charge, i.e. 
\begin{equation}
\label{eq9}
\partial I_3/\partial Q\geq 0,
\end{equation}
\begin{figure}[t]
\hspace{1.5cm}\includegraphics[width=0.50\textwidth]{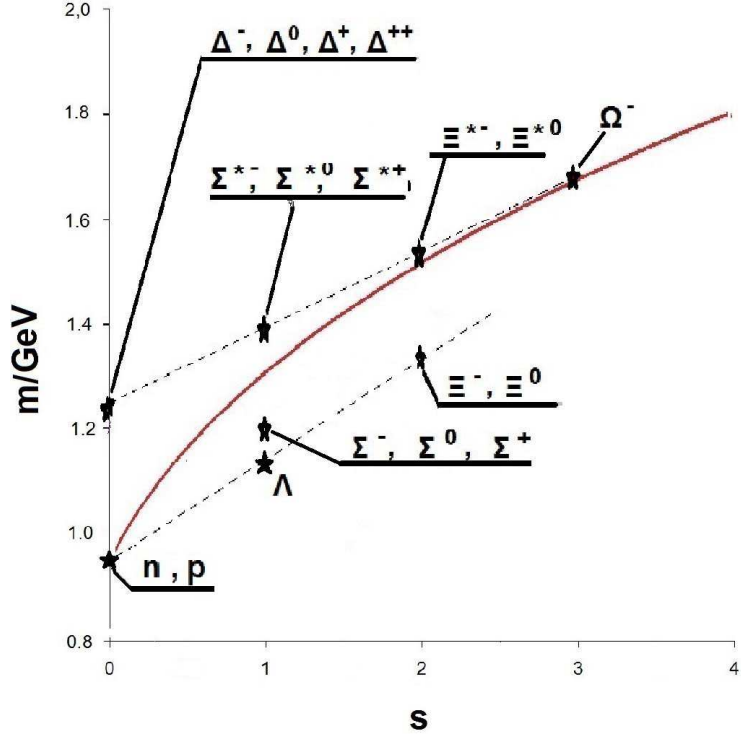}
\caption{Dependence of baryon mass on strangeness $S$.}
\label{fig:4}
\end{figure}
which is consistent with the Gellmann-Nishijima formula \cite{Nishijima55,Gell-Mann56}
\begin{equation}
\label{eq10}
Q=I_3+(1/2)(A+S),
\end{equation}
where $A$ is the baryon number and $S$ is the strangeness. The maximum $I_3$ value in a multiplet, $I_{3,max}$, is related to the number of members of a multiplet, $N$, via
\begin{equation}
\label{eq11}
I_{3,max}=\frac{N-1}{2}.
\end{equation}

Strangeness $(S)$ is attributed to the number of $S$ (strange) quarks in the hadron. As shown in Figures 1 and 2, strangeness increases with baryon mass. This is shown more clearly in Figure 4 which depicts the dependence of baryon masses on strangeness. Indeed $m$ increases with $S$ but there is considerabe scattering in the data and it is clear that $S$ alone cannot provide a satisfactory fit to the baryon masses. It appears from Fig. 4 that at least a second parameter is needed to provide a good description of the mass spectrum.

In fact, Figure 4 already shows that there appear to exist two families of baryons, one starting from $S=0$ with the proton and the neutron, and the other starting from $S=0$ with the $\Delta$ baryons and containing the $\Delta$, $\Sigma^*$, $\Xi^*$ and $\Omega^-$ baryons.

\section {Rotating neutrino model taxonomy}
\subsection{Experimental results}
\begin{table}[t]
\caption{Computed baryon masses via (eq. \ref{eq12}), i.e. $m=\left[n_{B}^{2} (2\ell +1)\right]_{}^{1/6} m_{p}$  and experimental values of baryon masses, the latter shown in parenthesis.} 
\label{tab:1}
\begin{center}
\begin{tabular}{p{0.4in}p{0.4in}p{0.7in}p{0.9in}p{0.4in}p{0.9in}p{0.5in}} \hline 
$n_{B}$ & $\ell_B$ & $m$ & $Particle$ & $S$ & $I_3$ & $Spin$ \\ \hline \hline
1 & 0 & 939 & p,n & 0 & 1/2, -1/2 & 1/2 \\ \hline 
1 & 1 & 1127 & $\Lambda$\newline (1116) & 1 & 1/2 & 1/2 \\ \hline 
1 & 2 & 1228 & $\Delta$\newline (1232) & 0 &-3/2,-1/2, 1/2,3/2& 3/2 \\ \hline 
1 & 3 & 1300 & \textbf{$\begin{array}{c} {\Xi ^{-} ,\Xi ^{o} } \\ {(1318)} \end{array}$} & 2 & 1/2, -1/2& 1/2 \\ \hline 
1 & 4 & 1356 & \textbf{$\begin{array}{c} {\Sigma ^{*} } \\ {(1384)} \end{array}$} & 1 & -1,0,1 & 3/2 \\ \hline \hline
2 & 0 & 1183 & \textbf{$\begin{array}{c} {\Sigma ^{-} ,\Sigma ^{o} ,\Sigma ^{+}} \\ {(1192)} \end{array}$} & 1 & -1,0,1 & 1/2 \\ \hline 
2 & 1 & 1420 & - & - & - \\ \hline 
2 & 2 & 1547 & \textbf{$\begin{array}{c} {\Xi ^{*,-} ,\Xi ^{*,o} } \\ {(1532)} \end{array}$} & 2 & -1/2,1/2 & 3/2 \\ \hline 
2 & 3 & 1636 & \textbf{$\begin{array}{c} {\Omega } \\ {(1672)} \end{array}$} & 3 & 0 & 3/2 \\ \hline\hline
\end{tabular}
\end{center}
\end{table}

\begin{figure}[ht]
\includegraphics[width=0.65\textwidth]{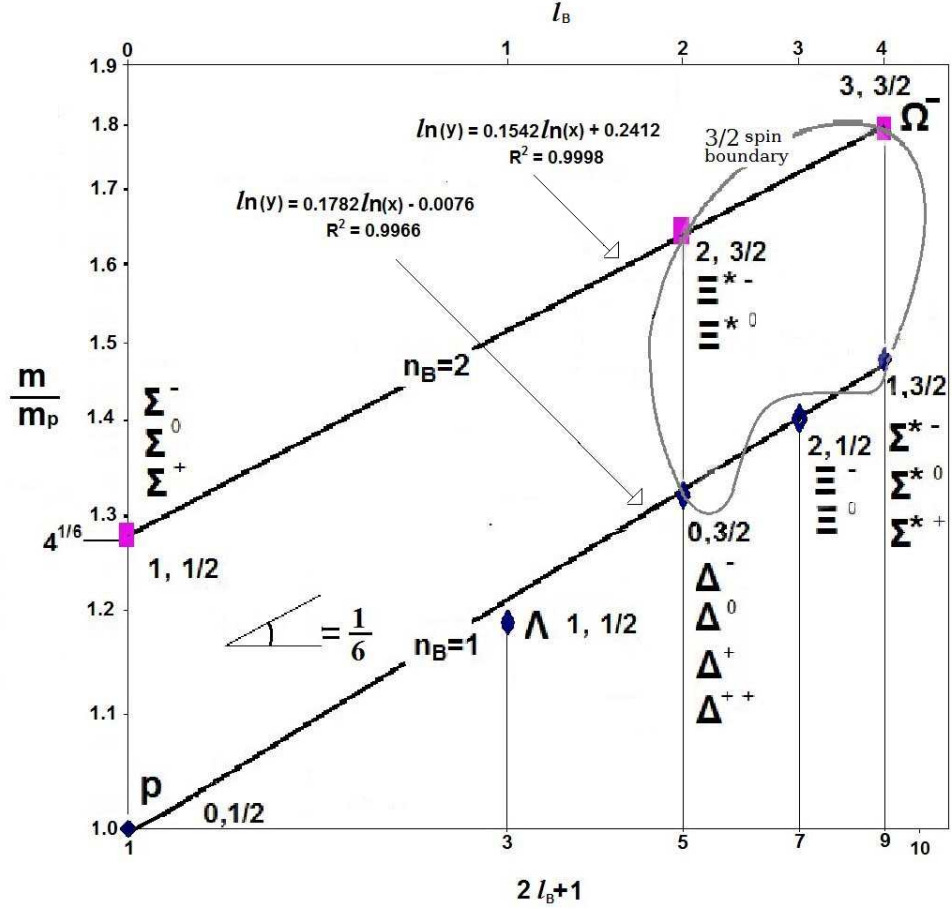}
\caption{Comparison of the masses, $m_B$, of the uncharmed baryons,
consisting of u, d and s quarks, with equation (\ref{eq12}), i.e. $m=m_p\left[n^2_B(2\ell_B+1)\right]^{1/6}$. Numbers next to each baryon denote strangeness and spin. With the exception of $\Xi^-$ and $\Xi^o$, which have spin 1/2, all baryons with $\ell\geq 2$ have spin 3/2.} 
\label{fig:5}
\end{figure}

As shown in Table 1 and in Figure 5, the masses of all uncharmed baryons, consisting of $u$, $d$ and $s$ quarks, fall, with a linear correlation coefficient $R^2$ better than 0.9966, on two parallel straight lines, corresponding to the equation
\begin{equation}
\label{eq12}
m=m_p\left[n^2_B(2\ell_B+1)\right]^{1/6},
\end{equation}
where $n_B$ is a positive integer (1,2,3,....) and $\ell_B$ is a non-negative integer (0,1,2,....).

In view of equation (\ref{eq5}) equation (\ref{eq12})  can also be written in the form
\begin{equation}
\label{eq13}
m=\left[n_B^2(2\ell_B+1)\right]^{1/6}3^{13/12}m^{1/3}_{Pl}m^{2/3}_o,
\end{equation}
where $m_p(=3^{13/12}m^{1/3}_{Pl}m_o^{2/3}=938.32$ MeV/c$^2$) is the proton mass.

The family $n_B=1$ includes, in addition to the proton and the neutron, the $\Lambda$, $\Delta$, $\Xi$ and $\Sigma^*$ baryons. On the other hand, the family $n_B=2$ contains the $\Sigma$, $\Xi^*$ and $\Omega$ baryons.

As it is discussed in the next subsection, the number $n_B$ bears some similarities with the principal quantum number, $n$, of the Bohr model for the H atom and to distinguish it here we use the subscript ``B'' (for Baryon). Also the second number $\ell_B$, may have some similarities with the second quantum number of the H atom Bohr model, thus we use the symbol $\ell$ with the same subscript ``B''.

Regarding the spin, we note that Figure 5 shows that all baryons with $\ell_B\geq 2$ have spin 3/2 (encircled area), with the exception of the $\Xi^-$ and $\Xi^o$ baryons which have spin 1/2. Also all baryons with $\ell <2$ have spin 1/2.

It should be noted that in previous works \cite{Vayenas12,Vayenas1106} we have attempted to describe the masses of the uncharmed baryons with a single quantum number, $n$, which in view of Figure 5 and Table 1, we replace here by the two quantum numbers, $n_B$ and $\ell_B$ which provide a quantitative fit to the mass spectrum (Fig. 5).

\subsection{Analogy with Bohr's H atom model}
In order to get some insight into the experimental results, which are described semiquantitatively by equations (\ref{eq12}) and (\ref{eq13}), it is useful to recall some key aspects of the Bohr model for the H atom. Bohr's original model published in 1913 \cite{Bohr1913} was based on the equation of motion for the electron, i.e. 
\begin{equation}
\label{eq14}
m_e\frac{\texttt{v}^2}{r}=\frac{e^2}{\varepsilon r^2},
\end{equation}
coupled with the assumption that the angular momentum, $L$, $(=m_e\texttt{v}r)$ is an integer multiple fo the Planck's constant $\hbar$, i.e. 
\begin{equation}
\label{eq15}
L=m_e\texttt{v} r=n\hbar,
\end{equation}
where $n$, now called the principle quantum number, is a positive integer. These two equations lead to the well known results
\begin{equation}
\label{eq16}
\texttt{v}=\alpha c/n\quad ; \quad \alpha=e^2/\varepsilon c\hbar,
\end{equation}
\begin{equation}
\label{eq17}
E_n=-\frac{m_e\alpha^2c^2}{2n^2}=-T=V/2,
\end{equation}
where $E_n$ are the energy levels of the electron and $T$ and $V$ denote its kinetic and potential energy respectively.

Some ten years later, after de Broglie published his famous de Broglie wavelength equation \cite{Broglie23}, i.e.
\begin{equation}
\label{eq18}
\lambda=\frac{h}{p}\quad ; \quad \mathchar'26\mkern-10mu\lambda=\frac{\hbar}{p}=\frac{\hbar}{m_e\texttt{v}},
\end{equation}
it became clear that equation (\ref{eq15}) can be interpreted as the ``standing wave condition'', namely that the electron is described by a wave where a whole number of wavelengths is required to fit along the circumference of the electron orbit. Indeed substituting the relation
\begin{equation}
\label{eq19}
2\pi r=n\lambda \quad;\quad r=n\mathchar'26\mkern-10mu\lambda,
\end{equation}
into equation (\ref{eq18}) and accounting for the definition of $L$ it follows that
\begin{equation}
\label{eq20}
L=m\texttt{v} r=m_e\texttt{v}\mathchar'26\mkern-10mu\lambda n=n\hbar,
\end{equation}
which is Bohr's assumption of Eq. (\ref{eq15}).

\subsection{Interpretation of the rotating neutrino taxonomy}
According to the rotating neutrino model \cite{Vayenas12,Vayenas1106}, a hadron consists of two building elements:
\begin{description}
	\item[a.] A rotating neutrino ring consisting of three (in the case of baryons) or two (in the case of mesons) relativistic neutrinos or antineutrinos held in orbit by the relativistic gravitational force. This bound rotational state is found here to be characterized by an integer number, $n^2_B(\ell_B+1)$, consisting of two integers $n_B$ and $\ell_B$. These integers may be viewed as two quantum numbers which bear some similarity with the principal quantum number, $n$, of the Bohr model for the H atom \cite{Vayenas16}. The fact that in the present case two, rather than one, principal quantum numbers are need to define the ground state may be related to the Pauli exclusion principle which does not allow more than two particles (e.g. neutrinos) to occupy the same state.
		\item[b.] A central mass consisting of one or rarely two leptons. These lepton(s) can be charged (e.g. $e^+$ in the case of a proton or $e^-$ in the case of the $\Sigma^-$ or $\Xi^-$ baryons), and are stabilized at the center of the rotating ring via ion-induced dipole or induced dipole-induced dipole electrostatic interactions \cite{Vayenas16,ICMSQUARE}.
\end{description}

Denoting by $\mathcal{L}_+$ and $\mathcal{L}_-$ the numbers of positively and negatively charged leptons at the central position, it follows that the charge $Q$ is given by $Q=\mathcal{L}_+-\mathcal{L}_-\equiv \mathcal{L}$. Equation (\ref{eq8}) implies $\partial I_3/\partial \mathcal{L}\geq 0$  and this suggests that the isospin may be related to the number of positively charged leptons at the center of the rotating neutrino ring. In fact from Figure 2 it follows that if $S=0$, then $I_3=\mathcal{L}-1/2$.

As already noted, according to the rotating neutrino model each of the three rotating neutrinos (Figs 3 and 6) is kept in orbit of radius $r$ centered at the center of gravity of the three neutrinos (denoted by O in Fig. 6), by the force expressed via the relativistic gravitational law, i.e. 
\begin{equation}
\label{eq21}
F=\gamma m_o\frac{\texttt{v}^2}{r}=\frac{Gm^2_o\gamma^6}{\sqrt{3}r^2},
\end{equation}
where $\gamma$ is the Lorentz factor $(1-\texttt{v}^2/c^2)^{-1/2}$, the factor $\sqrt{3}$ originates from the triangular geometry and $\gamma^3m_o$ is the inertial mass of each neutrino.

Noting that
\begin{equation}
\label{eq22}
\gamma m_o\frac{\texttt{v}^2}{r}=\frac{L\texttt{v}}{r^2},
\end{equation}
and that the angular momentum, $L$, equals $\gamma m_or \texttt{v}$ it follows that equation (\ref{eq21}) implies
\begin{equation}
\label{eq23}
\sqrt{3} L\texttt{v}=G m^2_o\gamma^6.
\end{equation}

This equation yields equation (\ref{eq12}) provided that $L$ is given by 
\begin{equation}
\label{eq24n}
L=n^2_B(2\ell_B+1)\hbar.
\end{equation}

This equation again can be interpreted as a standing wave condition where now instead of equation (\ref{eq19}) we have 
\begin{equation}
\label{eq24}
r=n^2_B(2\ell_B+1)\mathchar'26\mkern-10mu\lambda,
\end{equation}
where $n_B$ is a positive integer $(n_B=1,2,3,...)$  and $\ell_B$ is a non-negative integer (Fig. 6).

\begin{figure}[ht]
\begin{center}
\includegraphics[width=0.55\textwidth]{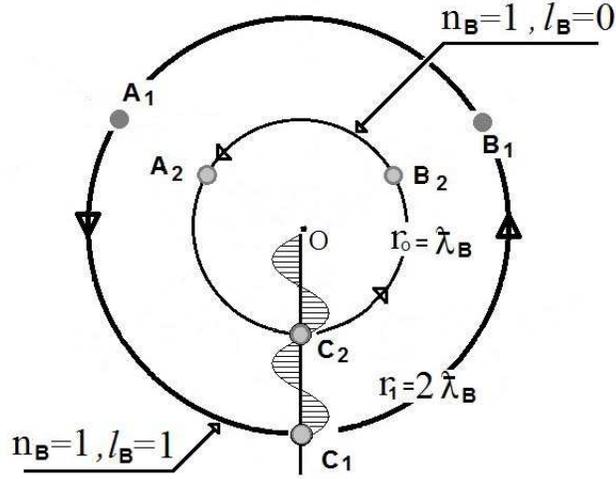}
\end{center}
\caption{Rotating neutrino model of baryon showing the Bohr-type  description of the neutrino orbits in terms of the two quantum numbers $n_B$ and $\ell_B$.}
\label{fig:6}
\end{figure}

Indeed substituting equation (\ref{eq24}) into the de Broglie wavelength equation, i.e. 
\begin{equation}
\label{eq25}
\mathchar'26\mkern-10mu\lambda=\frac{\hbar}{\gamma m_o\texttt{v}}
\end{equation}
and using the definition of the angular momentum $L$, we find equation (\ref{eq24n})
\begin{equation}
\label{eq26}
L=\gamma m_or \texttt{v}=n^2_B(2\ell_B+1)\hbar.
\end{equation}

As already noted the integer $\ell_B$ is related via $2\ell_B +1=2n-1$ to the positive integer $n$ $(n=1,2,....)$ which we have used in previous studies of the rotating neutrino model \cite{Vayenas12,Vayenas1106}.

If $\gamma >>1$, then $\texttt{v}\approx c$, thus replacing in (\ref{eq23}) $\texttt{v}$ by $c$ and $L$ by (\ref{eq24n}) we find
\begin{equation}
\label{eq28}
\sqrt{3}n^2_B(2\ell_B+1)\hbar c=Gm^2_o\gamma^6.
\end{equation}
Thus
\begin{equation}
\label{eq29}
\gamma=\left[\sqrt{3} n_B^2(2\ell_B+1)\right]^{1/6}\left(\frac{m_{Pl}}{m_o}\right)^{1/3},
\end{equation}

where $m_{Pl}(=\hbar c/G)^{1/2}$ is the Planck mass. Hence
\begin{equation}
\label{eq30}
m=3\gamma m_o=\left[n_B^2(2\ell_B+1)\right]^{1/6}3^{13/12}m^{1/3}_{Pl}m^{2/3}_o.
\end{equation}

The excellent agreement between equation (\ref{eq30}) and the experimental baryon mass spectrum is shown in Table 1 and figure 5.

\section{Conclusions}
In summary, the three quantum numbers of the SU(3) symmetry group of the Standard Model, namely charge, $Q$, Isospin, $I_3$, and strangeness, $S$, are analogous to the three quantum numbers, ($Q,n_B$ and $\mathchar'26\mkern-10mu\lambda_B$), which emerge from the rotating neutrino model. The baryon charge is determined by the charge, $\mathcal{L}$, of the positron(s) or electron(s) located at the center of the rotating neutrino ring. The maximum isospin value is also closely related to the number of leptons, $\mathcal{L}_+$ and $\mathcal{L}_-$,  which can be accomodated at the center of the rotating ring. 

Strangeness, on the other hand, is closely related both to the first quantum number, $n_B$, and to the second quantum number, $\ell_B$, which both dictate the value of the angular momentum and thus, via the Lorentz factor $\gamma$, the mass of the rotating neutrino baryon state given by
\begin{equation*}
m=\left[n^2_B(2\ell_B+1)\right]^{1/6}3^{13/12}m^{2/3}_{Pl}m^{1/3}_o.
\end{equation*}

This expression provides a very good fit ($R^2>0.996$) to the masses of uncharmed baryons. Thus, in analogy with Bohr's model of the H atom, it may turn out that the corresponding expressions for the Hamiltonians of these baryons, namely \cite{Vayenas12,Vayenas1106}, $-(2/3)m_p\left[n^2_B(2\ell_B+1)\right]^{1/6}c^2$ obtained via the rotating model, may be found in the future to be in reasonable agreement with the experimental values. However, the energies associated with transitions between such states are very high, in the far $\gamma-ray$ range, and thus may be difficult to obtain.

\end{document}